\documentclass{PoS}

\usepackage{graphicx}
\usepackage{amsmath}

\newcommand{\bee}{\begin{equation}}
\newcommand{\ee}{\end{equation}}
\newcommand{\beea}{\begin{eqnarray}}
\newcommand{\eea}{\end{eqnarray}}

\def\svev#1{\left\langle #1\right\rangle}       

\def\Kth{\kappa_\text{conf}}
\def\Kc{\kappa_c}

\title{Exploring the phase diagram of sextet QCD }

\ShortTitle{Exploring the phase diagram of sextet QCD }

\author{\speaker{Thomas DeGrand}\\
Physics Department,
        University of Colorado,
Boulder CO 80309 USA\\
        E-mail: \email{degrand@pizero.colorado.edu}}

\author{Yigal Shamir, Benjamin Svetitsky\\
Raymond and Beverly Sackler School of Physics and Astronomy,  Tel~Aviv
University, 69978 Tel~Aviv, Israel\\
        E-mail: \email{shamir@post.tau.ac.il}, \email{bqs@julian.tau.ac.il}}

\abstract{
As a follow up to the previous talk about the beta function of  $SU(3)$ gauge theory with
  $N_f=2$ symmetric representation (clover) fermions,
 we describe our explorations of the
beta-kappa plane, away from the massless limit.
Our simulations are mostly done on lattices of length $L=8$ and 12.
 We observe a phase transition from a strong coupling confined phase to
a deconfined, chirally restored phase. The line of transitions avoids (so far)
the location of the infrared fixed point discussed in the last talk.}

\FullConference{The XXVI International Symposium on Lattice Field Theory\\
		 July 14-19 2008\\
		 Williamsburg, Virginia, USA}

\begin{document}

This conference saw considerable discussion of  strongly interacting models which are
candidates for  physics beyond the Standard Model, which might be accessible at the LHC.
Before this year, most theoretical studies of these models were based on analytic techniques.
However, their Lagrangians are not that different from those of theories which lattice people
have been studying for years (like QCD). Perhaps lattice simulations can give new insights into
the dynamics of these models.

The Lagrangian we chose to study was an $SU(3)$ gauge theory coupled to two flavors of fermions
in the sextet (symmetric) representation.
 Ref.~\cite{Shamir:2008pb}, summarized
 in the companion presentation in this conference \cite{Ben}, described a Schrodinger functional study
of the running coupling constant. We observed behavior consistent with an
 infrared attractive fixed point (IRFP).
 This kind of study does not give direct evidence
of the spectrum and low energy constants of the system. That is the goal of the project we now
 describe.

Our lattice theory is defined by the single-plaquette gauge action and a
Wilson fermion action with added clover term~\cite{Sheikholeslami:1985ij}.
We  modify the clover term's coefficient via  tadpole improvement, $c_{SW}=1/u_0^3$.
A table of the relevant values is given in Ref.~\cite{Shamir:2008pb}.
All simulations used the standard hybrid Monte Carlo algorithm. The trajectories in various
runs were of lengths between 0.5 and 1.0, and the time steps ranged from~0.02 (at heavy quark
masses) to~0.005 (for light masses).
The data sets at each of our $(\beta,\kappa)$ values consist of 300 to 1000 trajectories,
with every fifth trajectory used for spectroscopy.

We want to map out the phase structure of the theory. We did this with a combination of
simulations on large lattice volumes, plus simulations in which one (or more) lattice directions
were small. When these directions become roughly the same size as the scale for some
physical process, they will affect it and give results which differ from what is seen on the large lattice.
Our picture for this description is a finite temperature transition, seen when the size of
temporal direction (in which the fermions obey antiperiodic boundary conditions) is smaller
than the sizes of the other directions. So we simulated volumes
\begin{itemize}
\item $8^4$, for quick scans
\item $12^3\times 8$, to reveal the critical $(\beta,\Kth)$ line where deconfinement occurs
(the last dimension is the temporal one)
\item $(12\times 8^2)\times 8$ allows faster runs than $12^3\times8$ and shows the same 
finite-temperature physics, though transitions are rounded by the smaller spatial volume.
\item $8^3\times 12$ is a ``zero temperature'' lattice compared to $N_t=8$.
We use it to study how the {\em spatial\/} size $L=8$ intrudes 
on the $q\bar q$ potential and on meson masses.
\item $12^4$ has two roles.  One is as a ``zero temperature'' lattice, which we use to determine 
zero-temperature quantities as long as all scales are short enough.
  The other role is as a rough  finite-temperature lattice that permits us to
 observe directly the movement of transition curves when $N_t$ changes from 8 to~12.
\end{itemize}
Dimensions of size 12 are where we determine meson masses.  
If the dimension is temporal, then the masses are conventional spectroscopic masses; 
if spatial, then the masses are screening masses, affected by the Matsubara
 frequencies that create non-zero momentum transverse to the meson propagation.
Lattices where $N_t=12$ are also where we calculate the (spatial)
 $q\bar q$ potential from Wilson loops. We used the ``P+A trick'' 
\cite{Blum:2001xb,Aoki:2005ga,Allton:2007hx,Aubin:2007pt}
 to construct propagators for
spectroscopy on our short lattices.

\begin{figure}
\begin{center}
\includegraphics[width=0.6\textwidth,clip]{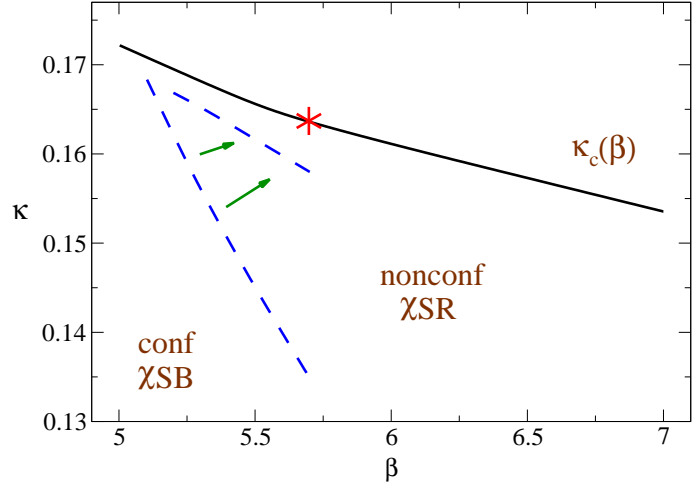}
\end{center}
\caption{Phase diagram  in the $(\beta,\kappa)$ plane.  The solid curve is
$\Kc(\beta)$, where $m_q$ vanishes; the dashed curves are
$\Kth(\beta)$, the confinement-deconfinement transition for $L=8$ (the curve to the left) and $L=12$
(the curve to the right).
The star on the $\Kc$ curve marks the approximate location of the IR fixed point found in
 Ref.~\cite{Shamir:2008pb}.
\label{fig:phase1}}
\end{figure}

The usual ordering of the Polyakov loop tells us
the location of the deconfinement transition. Fig. \ref{fig:phase1} shows our result
for two values of $L$, 8 and 12.
One interesting feature of the transition is that the fermions align the Polyakov loop
into a vacuum which spontaneously breaks charge conjugation, ${\rm Arg} \ P = \pm 2\pi/3$.
This means that the transition is a real phase transition for all quark mass, not
just a crossover. This behavior is expected from strong-coupling arguments \cite{Ogilvie_private}.
\begin{figure}
\begin{center}
\includegraphics[width=0.8\textwidth,clip]{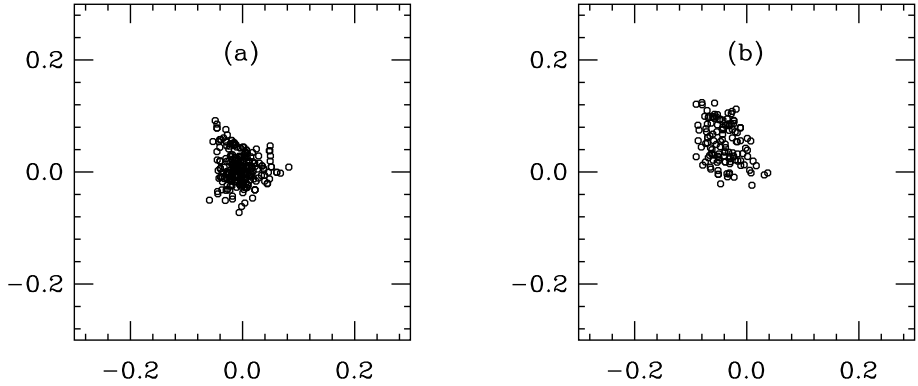}
\end{center}
\caption{Scatter plot of real and imaginary parts of the Polyakov loop from simulations at $\beta=5.7$, $12^3\times 8$ lattices.
(a) $\kappa=0.115$, in the confined phase (b) $\kappa=0.135$, in the deconfined phase.
\label{fig:scatter5.7}}
\end{figure}

 To describe the chiral properties of the theory
requires a little more work: we need a rather full set of observables, including the
condensate, the pseudoscalar decay constant, and masses of the pseudoscalar meson, as well as the masses of
the vector, axial vector, and scalar mesons.
It is hard to measure the condensate reliably for  Wilson fermions (without
doing something which amounts to invoking the GMOR relation).
It is both additively and multiplicatively renormalized. We have not used this observable.

A lattice determination of a continuum $f_\pi$ is a little involved. Because we are using
a non-chiral lattice action and a matrix element which does not precisely satisfy a Ward identity,
there is a lattice-to-continuum conversion factor $Z_A$  between the lattice matrix element,
the lattice spacing $a$, and a continuum-regulated decay constant $f_\pi$
\bee
af_\pi^{cont} = Z_A af_\pi^{latt}.
\ee
We are aware of two ways to compute $Z_A$. One is nonperturbative, through the RI (regularization
independent)
scheme, and the other is through perturbation theory. For our exploratory study, we believe perturbation
theory is adequate.
In the context of tadpole-improved perturbation theory \cite{Lepage:1992xa},
\bee Z_A = \left(1 + \frac{g^2}{16\pi^2} W\right)\left(1 - \frac{6\kappa}{8\kappa_c}\right) 
\ee
where $W$ is a numerical factor
 and the ``tadpole (TI) factor''
$\left(1 - \frac{6\kappa}{8\kappa_c}\right)$ corrects the
lattice field renormalization of lattice Wilson fermions compared to the continuum case.
We include the
tadpole factor in our
displayed results, but not the perturbative part of $Z_A$, which is independent of quark mass.

If chiral symmetry is broken, we expect to see $f_\pi$ extrapolating to a nonzero value in the 
chiral limit. If chiral symmetry is restored, $f_\pi$  will fall to zero.
In the chirally broken phase, we expect to see $m_\pi^2\sim m_q$. This is not the same
as observing $(am_\pi)^2\sim (am_q)$ along some arbitrary line in the space of bare parameters,
because the lattice spacing itself will change as the bare parameters change.
A more modest expectation is that at small quark mass the pion mass will become small compared to
all other massive quantities,  and that all these quantities, including $f_\pi$, will
remain nonzero in the chiral limit.

In a chirally restored phase we do not expect to see this mass hierarchy. Instead,
we expect to see parity doubling: the pseudoscalar and scalar ($a_0$)
 mesons should become degenerate,
as well as the vector and axial vector ($a_1$) mesons. This effect is seen in 
ordinary QCD \cite{Born:1991zz}, where near degeneracy of the (scalar, pseudoscalar) and (vector, 
axial vector)
multiplets is also observed. A naive expectation for a screening mass is that it
behaves  something like
\bee
m_H^2 = 4\left [ \left(\frac{\pi}{N_t}\right)^2 + m_q^2\right]
\ee
where $\frac{\pi}{N_t}$ is the lowest nonzero Matsubara frequency associated with 
antiperiodic boundary conditions in a lattice of temporal length $N_t$.

To complete the story, we  replace $\kappa$ by the Axial Ward Identity
quark mass, defined through
\bee
\partial_t \sum_x \svev{A_0(x,t)X(0)} = 2m_q \sum_x \svev{ P(x,t)X(0)}.
\label{AWI}
\ee
where $A_0=\bar \psi \gamma_0\gamma_5 \psi$ and $P = \bar \psi \gamma_5 \psi$.
  We neglect renormalizations and subtractions here as we did for $f_\pi$.
For consistency with the Schr\"odinger functional conventions, the derivative
is taken to be the naive nearest-neighbor difference.

We augment our observation of string tension and Polyakov loop with measurements of these
observables. In the confined phase, $am_\pi$, $am_\rho$, and $af_\pi$ are easy to
extract, while the $a_0$ and $a_1$ signals are poor. The  masses of the latter particles are large and
the fits are 
unstable. As we move into the deconfined phase, the signals in these channels improve
and the masses fall.

Everything we see in the deconfined phase is consistent with a picture of chiral
symmetry restoration. $af_\pi$ drops smoothly to zero as $am_q$ vanishes.
The pseudoscalar and vector meson screening masses remain nearly degenerate as the quark
mass is varied, and the $a_1$ and $a_0$ masses approach degeneracy with them.
The simplest explanation of what we see is that there is a single transition line,
at which confinement is lost and chiral symmetry is restored.

We show some plots which illustrate this behavior.
In all cases the right panel displays a massive quantity (in lattice units, $am_H$)
 versus $am_q$, while the
left panel displays the squared quantity  versus $am_q$. In all graphs,  crosses show the 
decay constant (with TI factor),
 diamonds the pion mass, squares the rho, octagons the $a_1$, and  bursts the $a_0$.

\begin{figure}
\begin{center}
\includegraphics[width=0.8\textwidth,clip]{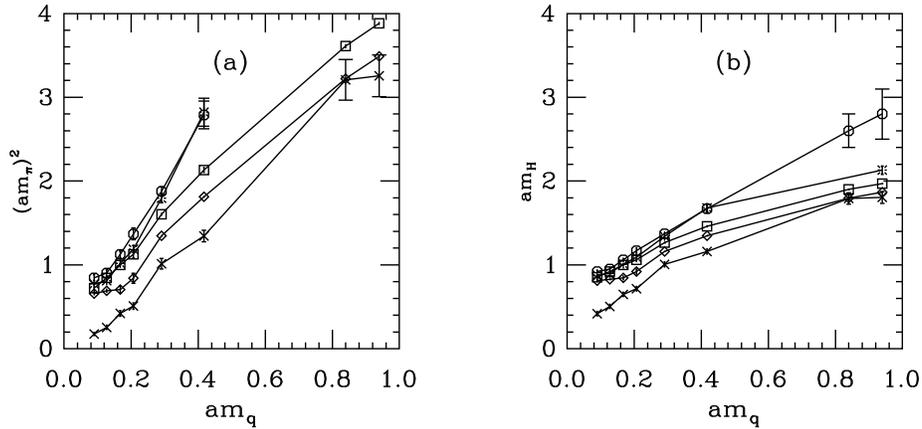}
\end{center}
\caption{ Spectroscopy for $\beta=5.5$ on $(12\times 8^2)\times 8$ volumes.
Crosses show $f_\pi$, pions are diamonds, rhos are squares, axial vector mesons are
 octagons are  and scalar mesons are bursts.
\label{fig:mpi5.51288}}
\end{figure}

We begin with a $(12\times 8^2)\times 8$ volume at $\beta=5.5$,
Fig. \ref{fig:mpi5.51288}. The two large quark mass points are confined.
The other points are deconfined. Although $(am_\pi)^2$ appears to vary linearly with $am_q$
down to small $am_q$,
none of the other criteria for chiral symmetry are satisfied. Instead, 
all states become degenerate and $af_\pi$ becomes small.

\begin{figure}
\begin{center}
\includegraphics[width=0.8\textwidth,clip]{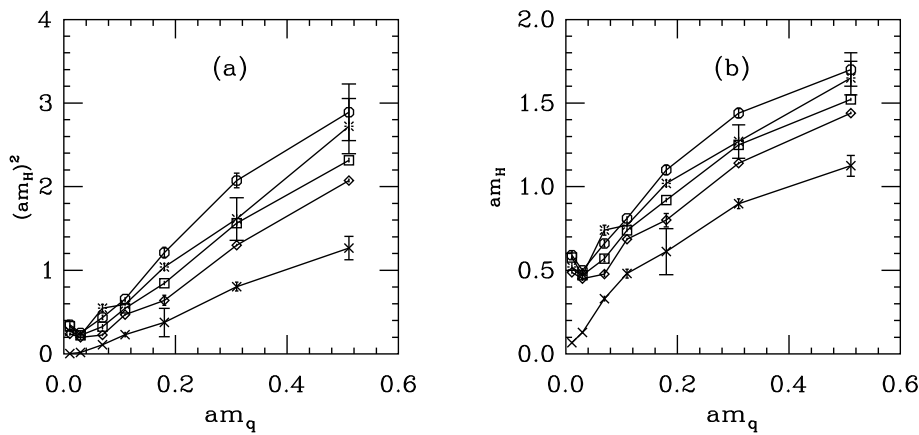}
\end{center}
\caption{ Spectroscopy for $\beta=5.7$ on $12^4$ volumes.
Crosses show $f_\pi$, pions are diamonds, rhos are squares, axial vector mesons are
 octagons are  and scalar mesons are bursts.
\label{fig:mpi5.71212}}
\end{figure}

We have also made rough observations of  the deconfinement line on our $12^4$ lattices,
away from the $\Kc$ line. It moves to larger $\beta$. Scans of spectroscopy show
similar behavior to what we saw on the smaller lattices. An example is shown in 
Fig. \ref{fig:mpi5.71212}, $\beta=5.7$:
The two heaviest mass points are confined.
The next point is on the crossover and the rest are deconfined.
These are not screening masses, but measurements performed in the temporal direction (ordinary
spectroscopy). We have checked screening correlators at several of these points and they 
produce identical results.

The two-phase structure with a single transition line is reminiscent of what is seen in QCD
 with a small
number of flavors of fundamental-representation fermions.
 The motion of the deconfinement transition line at larger quark mass is to be expected. 
At larger
quark mass (at smaller $\kappa$) the quarks decouple and our theory reduces to a pure gauge
system. At increasing $N_t$ the boundary between confinement-like behavior
and deconfinement will move to larger $\beta$ and its flow along lines of 
constant physics (for example, fixed large quark mass)  will be governed by
 asymptotic scaling for $T_c$. The behavior
of the deconfinement line near $\kappa_c$ will be different if the zero temperature theory has an 
infrared fixed point, than if it is everywhere confining, like QCD. In the latter
case the transition line will also move out the $\Kc$ line to ever larger $\beta$.
But a theory with an IRFP must have a basin of attraction in which the gauge coupling flows
into the FP. In this basin, the theory is deconfined and chirally restored, so no transition
(or line of transitions) to a confined, chirally broken phase can intrude. 
Thus the intersection point of the deconfinement curve with $\kappa_c$ must come to a halt
at sufficiently large $N_t$.
Zero temperature simulations might identify this point with a bulk transition point,
which must be present at zero quark mass to separate the (confining) strongly coupled
phase from the IRFP's basin of attraction.

We  notice qualitative similarities between our system and observations of $SU(2)$ gauge theory
 with $N_f=2$ flavors
of adjoint fermions, reported at this meeting or near
 it \cite{Catterall:2008qk,Hietanen,DelDebbio:2008zf}.
In these systems, at large $\beta$ and near $\Kc$ the $\pi/\rho$ mass ratio appears
to remain near unity while the string tension $a^2\sigma$ becomes small.
As far as we know, none of these groups have done finite temperature simulations as a diagnostic
for the location of phase boundaries, nor have they compared $f_\pi$, or the states whose masses 
which might parity-double, over
a wide parameter range. It would be very useful to add these tests to those already in use, to
further characterize the phase structure of these theories.

$SU(2)$ pure gauge theories possess a second order confinement - deconfinement
transition and adjoint fermions preserve the $Z(2)$ center symmetry, so the pattern
of finite temperature transitions sweeping across the phase diagram is expected
to be present here, too.

Finally, we would like to remark: the description of our data that we  have presented here is
rather different from the one the speaker (T. D.) gave at the conference.
At that time we thought we were seeing a separate chiral symmetry restoration transition at weaker
coupling.
Such behavior would be a true (second order) transition only at $am_q=0$ and
 in the continuum limit, and
would only be a crossover at nonzero quark mass (where all our data is taken).
It also cannot approach the location of the IRFP.
The rough linearity of $(am_\pi)^2$ vs $am_q$, which describes our data through much of
the parameter space where we are deconfined, fooled us.
But, as we have described above, there is more to chiral symmetry breaking than one mass relation.

This work was supported in part by the US Department of Energy and by the Israel Science Foundation
 under grant
no.~173/05.  Our computer code is based on version 7 of the publicly available code of the MILC
collaboration~\cite{MILC}. We would like to thank   N.~Christ, A.~Hasenfratz,
 T.~G.~Kovacs, and M.~Ogilvie for discussions.

\end{document}